\documentclass[%
reprint,
amsmath,amssymb,
aps,
prb,
floatfix,
]{revtex4-2}

\usepackage{graphicx}%
\usepackage{dcolumn}%
\usepackage{bm}%
\usepackage{xcolor}%

\begin{document}

\preprint{APS/123-QED}

\title{Effect of the depolarizing field on the domain structure of an improper ferroelectric}%

\author{Aaron Merlin Müller}
 \affiliation{%
 Department of Materials, ETH Zurich, 8093 Zurich, Switzerland
}%
\author{Amadé Bortis}%
\affiliation{%
 Department of Materials, ETH Zurich, 8093 Zurich, Switzerland
}%

\author{Arkadiy Simonov}%
\affiliation{%
 Department of Materials, ETH Zurich, 8093 Zurich, Switzerland
}%

\author{Manfred Fiebig}%
\email{manfred.fiebig@mat.ethz.ch}
\affiliation{%
 Department of Materials, ETH Zurich, 8093 Zurich, Switzerland
}%

\author{Thomas Lottermoser}%
\affiliation{%
 Department of Materials, ETH Zurich, 8093 Zurich, Switzerland
}%

\date{\today}%

\begin{abstract}
We show that, contrary to common belief, the depolarizing electric field generated by bound charges at thin-film surfaces can have a substantial impact on the domain structure of an improper ferroelectric
with topological defects.
In hexagonal-manganite thin films, we observe in phase-field simulations that through the action of the depolarizing field, 
(i) the average magnitude of the polarization decreases, (ii) the local magnitude of the polarization decreases with increasing distance from the domain walls, and (iii) there is a significant alteration of the domain-size distribution and average domain size, which is visualized with the pair-correlation function.
We conclude that, in general, it is not appropriate to ignore the effects of the depolarizing field for thin film ferroelectrics.
\end{abstract}

\maketitle

\section{Introduction}

Thin-film ferroelectric materials are promising ingredients for a new generation of nanoelectronic devices, including memristors \cite{chanthbouala_ferroelectric_2012}, magnetoelectronic storage \cite{trassin_low_2015,heron_deterministic_2014}, and re-writable circuits of conducting domain walls \cite{meier_ferroelectric_2022}. All these functionalities critically depend on the distribution of ferroelectric domains. 
Here we distinguish between two classes of ferroelectric materials with fundamentally different types of domain formation and distribution. These are proper ferroelectrics, where the primary order of the system is the electric polarization, and improper ferroelectrics, where the primary order of the system is associated with magnetic or distortive order that drives the electric polarization as a secondary effect. 

For thin films of proper ferroelectrics, the depolarizing field, a static electic field that is generated by the bound charges at the surfaces and interfaces of the film, is a major driving force behind domain formation. Without the electrostatic effect of the depolarizing field, the minimal energy state for proper ferroelectric thin films would be a single-domain configuration.
This contrasts with the case of improper ferroelectrics, where the polar domain structure commonly follows the domain structure of the primary distortive or magnetic order \cite{artyukhin_landau_2014, fiebig_evolution_2016}.
A depolarizing field is also present, but
because of the dominant influence of the primary order, its influence is generally neglected.

In this work, we predict that even in improper ferroelectrics and despite the presence of topological defects pinning the domains 
the depolarizing field can have an unexpected, significant impact on the domain structure.
We perform phase-field simulations of free-standing thin films of hexagonal manganites, a lattice-distortively driven improper ferroelectric, and show that 
the average magnitude of the spontaneous electric polarization is lowered by the depolarizing field. Remarkably, the magnitude of the polarization drops with increasing distance from the domain walls. 
Furthermore, the Fourier-transformed intensities and the cross-correlation function reveal a significant effect on the domain-size distribution. Both change from a 2D-Gaussian function in the absence of a depolarizing field to donut-like distribution in its presence.
This demonstrates that it is not appropriate to ignore the depolarizing field in thin film improper ferroelectrics, regardless of topological defects.

\section{The Landau expansion of hexagonal manganites} \label{Landau}

\begin{figure}
    \centering
    \includegraphics[width=0.40\textwidth]{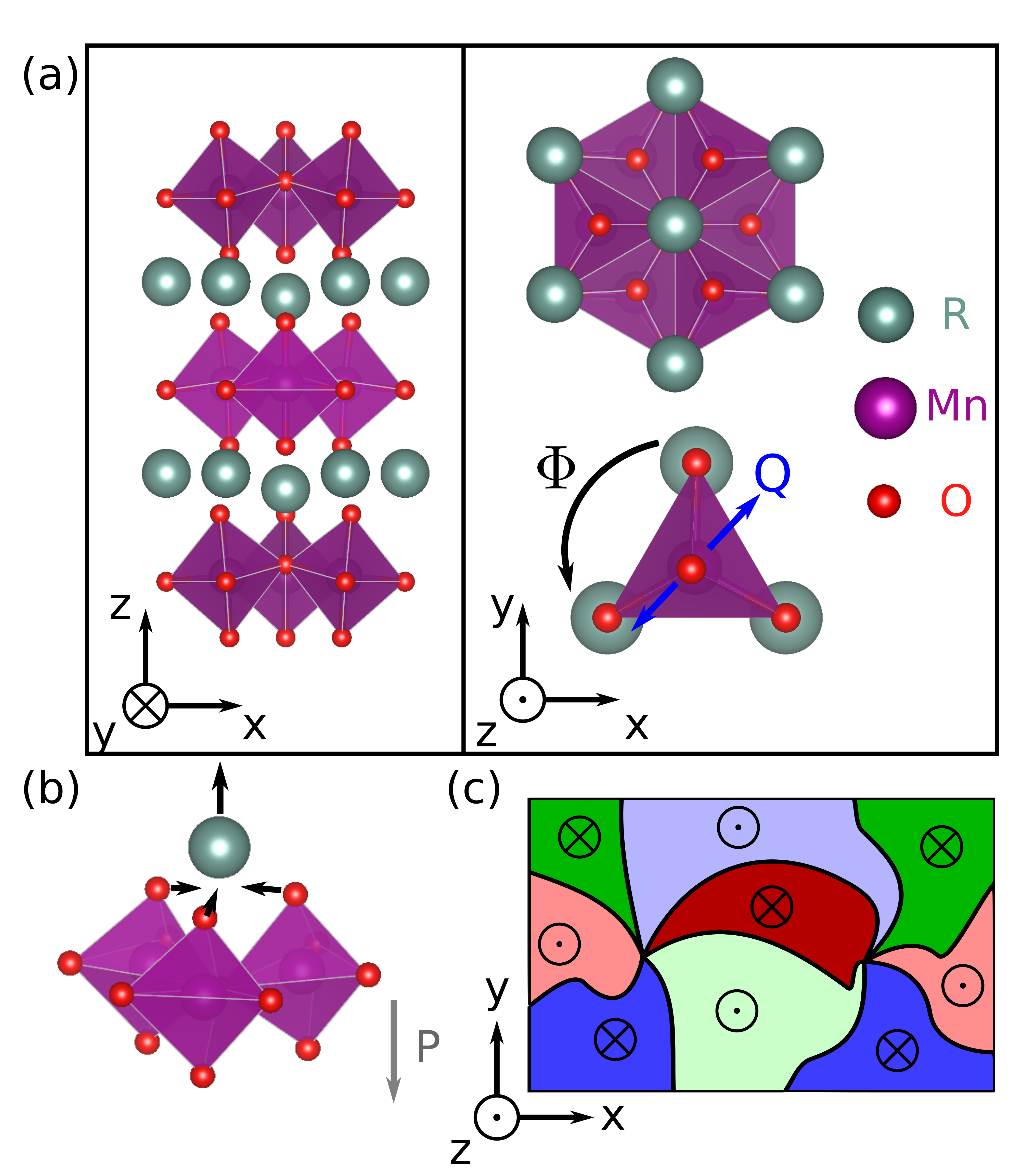}
    \caption{(a) Visualization of the structural order of hexagonal manganites. The order is given by a tilt of the MnO$_{5}$ bipyramids parametrized by the zone-boundary mode $\mathrm{K}_3$. The two-dimensional order parameter is illustrated in polar coordinates with $Q$ as amplitude and $\Phi$ as azimuthal angle of the MnO$_5$ tilt. Three neighboring bipyramids form a trimer and tilt towards a common center, which triples the unit cell \cite{sim_hexagonal_2016}. (b) Visualization of the deformation of the MnO$_5$ bipyramids during a transition from the paraelectric to the ferroelectric phase. Small asymmetries in this process cause polarization onset \cite{fennie_ferroelectric_2005}. (c) Schematic drawing of a conceptual distortive-ferroelectric domain structure. Topologically protected vortices are formed by the six domain states of the structural order. 
    The circular symbols denote the improper ferroelectric out-of-plane polarization which follows an alternating pattern. }
    \label{fig:hexmag_orders}
\end{figure}

 The spontaneous long-range order of hexagonal manganites is driven by an inverson-symmetry-breaking tilt of its MnO$_5$ bipyramids and is visualized in Fig.~\ref{fig:hexmag_orders}a. 
This primary order is described by a two-dimensional order parameter $\mathbf{Q}$, which corresponds to the zone-boundary mode $\mathrm{K}_3$ \cite{fennie_ferroelectric_2005}.
In polar coordinates, the radius corresponds to $Q = |\mathbf{Q}|$ and the azimuthal angle of the lattice-trimerizing bipyramidal MnO$_5$ tilt corresponds to $\Phi$.
This primary structural order is then coupled to a secondary ferroelectric order, which is described by a displacement field associated with a one-dimensional order parameter $\mathcal{P}$.
The displacement field corresponds to the polar mode $\Gamma^{-}_{2}$, and the order parameter $\mathcal{P}$ corresponds to the amplitude of the associated phonon, with a spontaneous polarization $P_\mathrm{s} \propto \mathcal{P}$.

The Landau free energy for this order is \cite{artyukhin_landau_2014}

\begin{widetext}
\begin{align} \notag
        F & = \frac{a}{2} Q^2 + \frac{b}{4} Q^4 + \frac{Q^6}{6}  ( c + c' \, \text{cos} \, 6 \Phi ) & \text{Structural Order} \\
        & - g \, Q^3 \,  \mathcal{P} \, \text{cos} 3 \Phi + \frac{g'}{2} Q^2 \mathcal{P}^2 + \frac{a_\mathcal{P}}{2} \, \mathcal{P}^2 &  \text{Improper Order} \label{eq:FreeEnergy}\\
        + \frac{1}{2} \sum \limits_{i = x,y,z} & [ s_Q^i ( \partial_i Q \partial_i Q + Q^2 \partial_i \Phi \partial_i \Phi ) + s_\mathcal{P}^i \, \partial_i \mathcal{P} \partial_i \mathcal{P} ]. & \text{Stiffness} \notag
\end{align}
\end{widetext}

The first line in Eq.~\ref{eq:FreeEnergy} corresponds to the Landau expansion of the primary structural order, and the second line corresponds to the coupling between the primary structural and the secondary ferroelectric order. The third line contains stiffness terms that result in energy penalties for domain walls (see Appendix~\ref{app:NumericalMethodsAppendix} for more details). A schematic domain pattern resulting from the Landau expansion in Eq.~\ref{eq:FreeEnergy} is visualized in Fig.~\ref{fig:hexmag_orders}c. There are six trimerization-polarization domain states and the associated domains form vortices of six domains in a sequence of these states and 
with an alternating polarization along the hexagonal axis.
The vortices are topological defects because they are robust to local perturbations of the domain structure and can only be created and annihilated in pairs \cite{artyukhin_landau_2014}.

To model the effect of the the electrostatic interaction, we include an additional term in Eq.~\ref{eq:FreeEnergy} \cite{bortis_manipulation_2022, smabraten_charged_2018, yang_dynamical_2022}, given by 

\begin{equation}
F_\text{electrostatic} = -P_\mathrm{s} E \; ,
\end{equation}

\noindent where the electric field $E$ is obtained from the Gauss law

\begin{equation}
    \nabla E = - \frac{\nabla P_\mathrm{s}}{\varepsilon_{\rm b} \varepsilon_0}.
\end{equation}

\noindent Here, $\varepsilon_0$ is the vacuum permittivity and $\varepsilon_{\rm b}$ is the background dielectric constant \cite{levanyukBackgroundDielectricPermittivity2016, tagantsevLandauExpansionFerroelectrics2008a}. The background dielectric constant contains the electronic dielectric response and the dielectric response of all normal modes in the system with the exception of the mode describing the polarization $P$. Here, we choose a background dielectric constant of $\varepsilon_{\rm b} = 1$ to make the electrostatic effects more evident. As shown in the supplemental material, increasing the background dielectric constant to the experimental value of $\varepsilon_{\rm b} = 8.9$ \cite{ruff_frequency_2018, smabraten_charged_2018} merely decreases the effect of the depolarizing field but does not affect the results qualitatively. On the practical side, choosing the value $\varepsilon_{\rm b} = 1$ permits us to reduce the size of our computational mesh while keeping the required computational time within feasible limits.

The model in Eq.~\ref{eq:FreeEnergy} is derived for bulk crystals. Here, we adapt it to the description of thin films by using periodic boundary conditions in the in-plane directions and open boundary conditions in the out-of-plane direction, as previously done in Ref. \cite{bortis_manipulation_2022}. Furthermore, we assume that the system is a perfect insulator and that it is devoid of any free charges. We choose open boundary conditions over closed-circuit boundary conditions to bring out the effects of the electrostatic interaction as clearly as possible and to avoid the significant complexity and computational time of simulating free charge carriers and charge currents \cite{yang_dynamical_2022}.

As mentioned, we use phase-field simulations to obtain the domain structure exhibited by the hexagonal manganites \cite{chen_phase-field_2002,wang_understanding_2019, xue_evolution_2015, xue_topological_2018}. In phase-field simulations, the mesoscopic order and domain pattern of the system are described as a continuous field of order parameters. This field is initialized with random values. The subsequent evolution of the system is given by the Ginzburg-Landau equation

\begin{equation}
    \frac{\partial \eta}{\partial t} = - \frac{\delta F}{\delta \eta},
\end{equation}

\noindent where $\eta$ is an order parameter, here $Q$, $\Phi$, or $\mathcal{P}$. The expression $\delta / \delta \eta$ corresponds to a functional derivative. The electrostatic field of the system is computed from the Gauss law \cite{smabraten_charged_2018}. Finally, we compute the Fourier-transformed intensity of the polarization-density field by taking the square of the Fourier transform of the system, in equivalence to scattering experiments. The Fourier-transformed intensity then gives us insight on the domain-size distribution. 
The pair correlation of the system is computed from an inverse transform of the intensity to derive typical domain sizes. A detailed description of the computational details and data analysis can be found in Appendix~\ref{app:NumericalMethodsAppendix}. As shown in the supplemental material, the full-width half-maximum of the central peak of the pair correlation is proportional to the average domain size \cite{giraldoMagnetoelectricDomainEngineering2024a}.

\section{Results}

\begin{figure}
    \centering
    \includegraphics[width=0.5\textwidth]{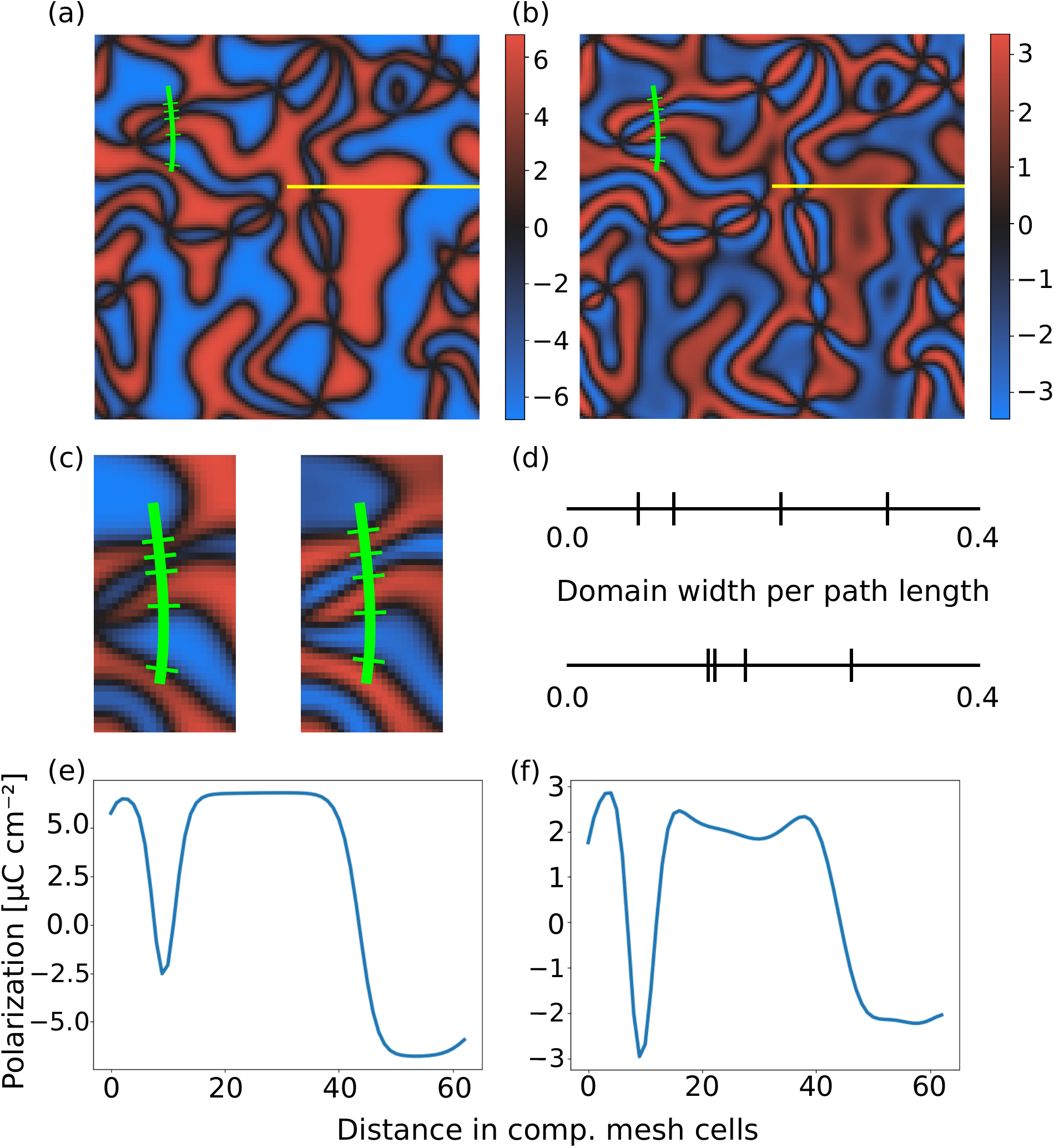}
    \caption{Dependence of improper-ferroelectric order on depolarizing field, visualized with phase-field simulations in real space. 
    (a) Domain pattern for a simulation that ignores the electrostatic interaction. (b) Domain pattern for a simulation that includes the electrostatic interaction assuming a background dielectric constant $\varepsilon_{\rm b} = 1$ (see text on this choice).
    (c) Close-up on the green paths through panels (a) (left) and (b) (right).
    (d) Domain size along the green path for the simulations without (top) and with (bottom) the depolarizing field.
    (e, f) Polarization profile along a section, shown as yellow line in (a, b). %
    The gradient and resulting dip in polarization is clearly visible in (f).
    }
    \label{fig:RealSpaceImages}
\end{figure}

\begin{figure}
    \centering
    \includegraphics[width=0.5\textwidth]{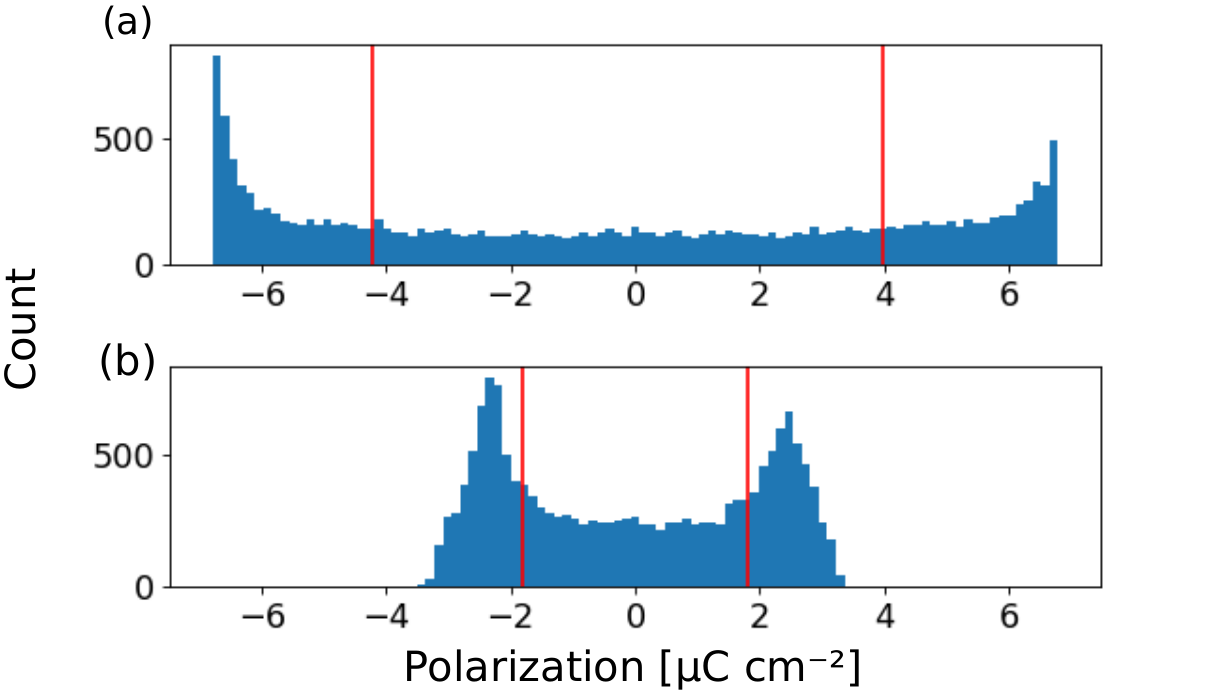}
    \caption{Histogram of the polarization per point in the computational mesh for a simulation (a) without and (b) with consideration of the electrostatic term. The mean of positive/negative values are shown with red vertical bars. For both histograms, the data of Fig.~\ref{fig:RealSpaceImages} have been used. The average magnitude of the polarization is lower in (b) compared to (a), which illustrates the influence of the depolarizing field.
    }
    \label{fig:Histograms}
\end{figure}

\begin{figure*}
    \centering
    \includegraphics[width=0.98\textwidth]{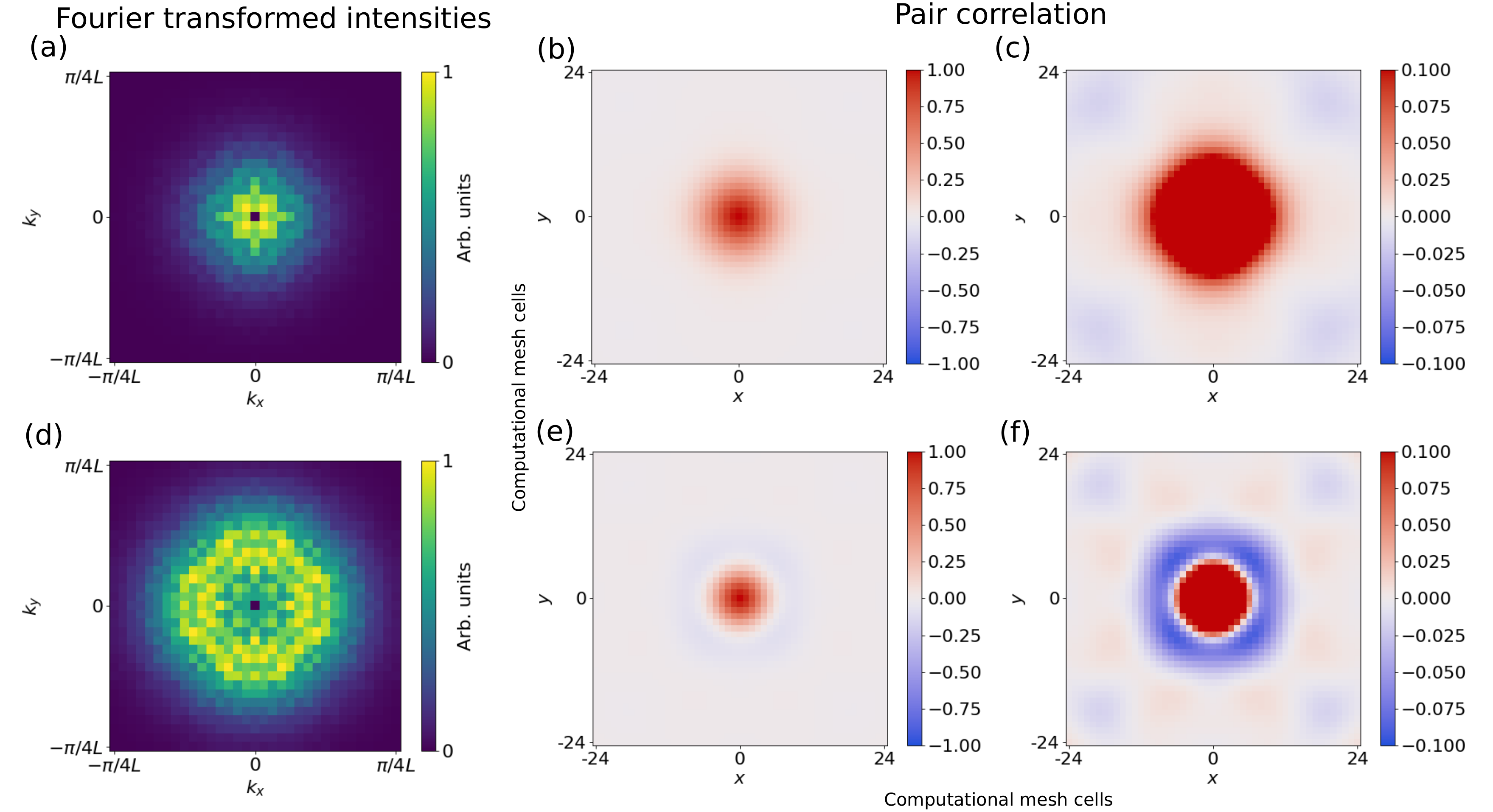}
    \caption{ Fourier-transformed intensity and pair correlation of twenty combined simulations for (a-c) systems without electrostatic interaction and (d-f) systems with  electrostatic interaction. (a) Fourier-transformed intensity and (b,c) pair correlation with two scaling ranges. (d-f) Same as (a-c), but with  electrostatic interaction for $\varepsilon_{\rm b} = 1$. Note that data augmentation was used for this analysis (see Appendix~\ref{app:NumericalMethodsAppendix}). The anisotropy of the plot results from the quadratic simulation mesh and the artifacts its size limitation introduces.     
    If electrostatic interaction and hence a depolarizing field is present, the ring-like intensity in (d) and the spherical wave in (e) and (f) indicate a narrowing of the domain-size distribution, whereas a smaller width of the central peak corresponds to a decrease in the average domain size.
    }
    \label{fig:PairCorrelation}
\end{figure*}

\begin{figure}
    \centering
    \includegraphics[width=0.5\textwidth]{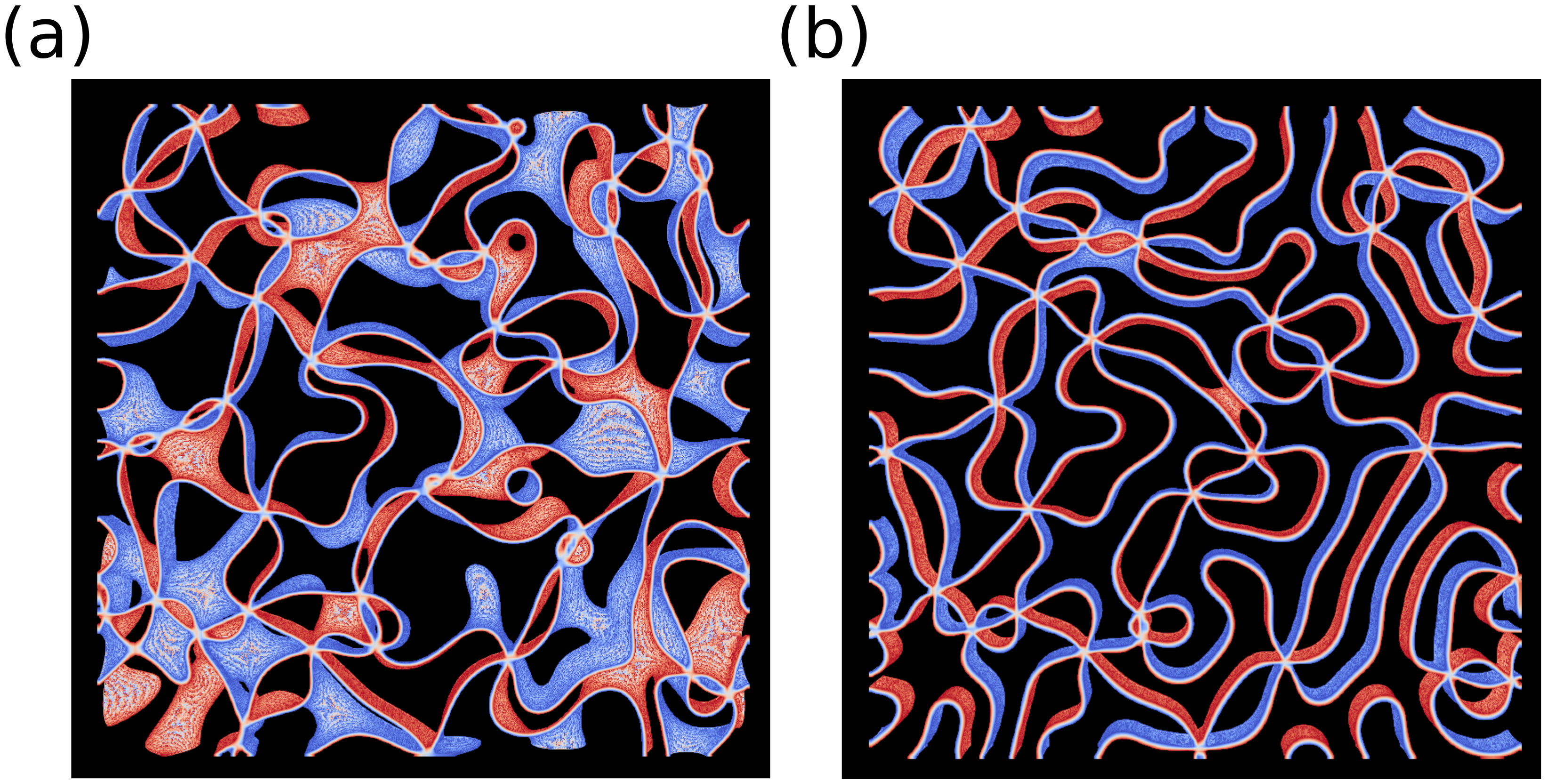}
    \caption{3D-distribution of domain walls in simulations of thin films with and without electrostatic interaction. The color signifies the magnitude of the polarization in the adjacent domain. (a) Simulation ignoring electrostatic interaction and (b) simulation including electrostatic interaction. 
    In (b), the additional electrostatic penalty for domain walls perpendicular to the z-direction cause a much stronger alignment of the domain walls along the z-axis than in (a).}

    \label{fig:ThickFilms}
\end{figure}

\begin{figure}
    \centering
    \includegraphics[width=0.5\textwidth]{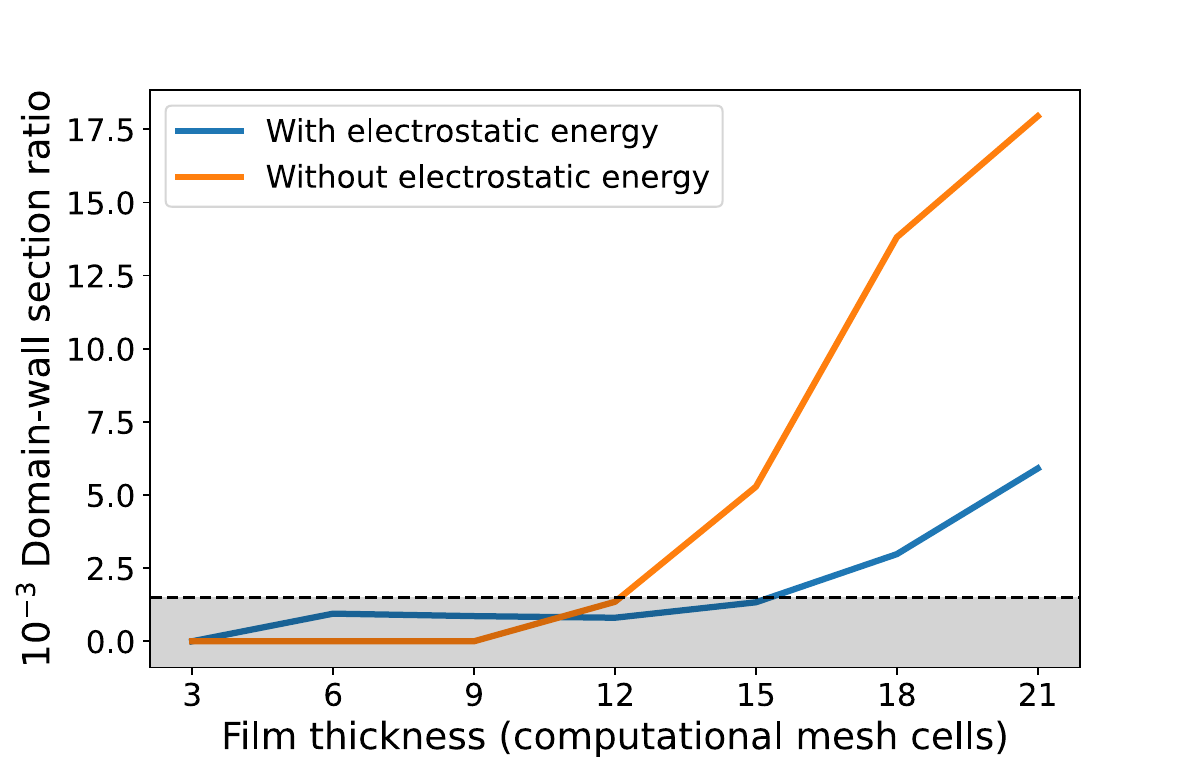}
    \caption{Fraction of domain-walls sections oriented perpendicular to the z-direction for simulations of thin films of different thickness.
    The electrostatic interaction suppresses this orientation, as shown above.
    An error (grey region) of about 1.5\% for thin film simulations with a low number of computational mesh cells stems from numerical errors of the Poisson equation solver.}
    \label{fig:ThickFilmsQuantitative}
\end{figure}

The electrostatic interaction can be divided into two contributions. Bound charges at the surface of the thin film give raise to the depolarizing field, which is distinguished from the electric field generated by bound charges at domain walls. We treat the effect of the depolarizing field in Figs.~\ref{fig:RealSpaceImages}-\ref{fig:PairCorrelation}. The effect of bound charges at domain walls is treated in Figs.~\ref{fig:ThickFilms} and \ref{fig:ThickFilmsQuantitative}.

Figure~\ref{fig:RealSpaceImages} shows domain patterns obtained by simulations with and without inclusion of the electrostatic interactions, both cases starting from the same initial configuration.
 In Figs.~\ref{fig:RealSpaceImages}-\ref{fig:PairCorrelation}, only the depolarizing field contributes to electrostatic interactions because the domain walls are parallel to the polarization and harbor no bound charges.
A number of differences are immediately visible. 
Without consideration of electrostatic effects, our simulation results in typical domain patterns of hexagonal manganites
\cite{giraldo_magnetoelectric_2021, artyukhin_landau_2014}. 
The magnitude of the polarization is constant all across the expansion of the domain outside the region of the domain walls \cite{xue_topological_2018,bortis_manipulation_2022}. 
In the simulation including the electrostatic interactions in Eq.~\ref{eq:FreeEnergy}, however, there is a decline in polarization that deepens the further the distance to the domain walls is.
This is further illustrated in the polarization profiles Figs.~\ref{fig:RealSpaceImages}e and \ref{fig:RealSpaceImages}f along the yellow cuts in Figs.~\ref{fig:RealSpaceImages}a and \ref{fig:RealSpaceImages}b.
This remarkable variation of the polarization distribution is a consequence of the non-local nature of the electrostatic interaction.
Close to a domain wall, the electrostatic contributions from the domains on the two sides of the domain wall cancel partially, so the net electrostatic field is weak, resulting in a higher polarization magnitude.
If, on the other hand, a point is far away from a domain wall, the electrostatic field is increased, resulting in the observed decline of polarization away from the walls with a dip in the approximate center of the domain.
In Fig.~\ref{fig:Histograms}, we present the distribution of the local polarization of our system. 
Comparing Figs.~\ref{fig:Histograms}a and \ref{fig:Histograms}b, one can see that the electrostatic interaction lowers the average magnitude of the polarization of the system, but does not completely suppress it.

In Fig.~\ref{fig:PairCorrelation}, the Fourier-transformed intensities and the pair correlation of the polarization has been computed by summing the result of twenty simulations with different initial fields. We further perform data augmentation. This computation is equivalent to the simulation of a scattering experiment, and its technical details are described in Appendix~\ref{app:NumericalMethodsAppendix}. 
Figures~\ref{fig:PairCorrelation}a, \ref{fig:PairCorrelation}b, and \ref{fig:PairCorrelation}c show the exponentially decaying intensity and pair correlation of a system that does not include the  electrostatic interaction. The intensity is indicative of a broad domain-size distribution that is typical for hexagonal manganites in bulk \cite{giraldo_magnetoelectric_2021} or for thin films where the depolarizing field is fully screened \cite{meier_global_2017, xue_topological_2018}. This contrasts strikingly with the intensity distribution in simulations that include the electrostatic interaction. The observed ring in Fig.~\ref{fig:PairCorrelation}d and the pair correlation in Figs.~\ref{fig:PairCorrelation}e and \ref{fig:PairCorrelation}f, which takes the form of a spherical wave, reveal a narrowing of the domain-size distribution, whereas a smaller width of the central peak corresponds to a decrease in the average domain size, an aspect that is further discussed in the supplemental material. 
This change in the domain patterns is further illustrated in real space in Fig.~\ref{fig:RealSpaceImages}c, which shows an illustrative path exemplifying the change in the shape of the domains, and in and in Fig.~\ref{fig:RealSpaceImages}d, which shows the associated change in domain-size distribution.

Finally, in Fig.~\ref{fig:ThickFilms} we show that the  electric field generated by bound charges at domain walls tends to align domain walls parallel to the z-axis. Figure~\ref{fig:ThickFilms}a shows a simulation of a film that is thicker than in the previous simulations to the extent that domain walls are no longer exclusively aligned parallel to the z-direction. Instead, they start to bend towards the xy-plane, which indicates a cross-over from the thin-film regime to the bulk regime.
If the electrostatic interaction is considered, as in Fig.~\ref{fig:ThickFilms}b, the alignment along $z$ is drastically enhanced and most domain-wall sections lie parallel to the z-direction.
On domain walls perpendicular to the z-direction, and hence perpendicular to the polarization, bound charges will accumulate at head-to-head and tail-to-tail domain walls.
These bound charges generate an electric field that increases the energy penalty for such domain walls significantly.

The influence of the electric field generated by bound charges at domain walls on the wall alignment is further illustrated in the quantitative analysis shown in Fig.~\ref{fig:ThickFilmsQuantitative}. We plot the fraction of domain-wall sections that are oriented perpendicular to the z-direction of systems with varying thickness for simulations with and without the electrostatic interaction (computational details in Appendix~\ref{app:NumericalMethodsAppendix}). In films that are very thin, all domain walls are aligned along the z-direction for both scenarios because the anisotropic stiffness terms in Eq.~\ref{eq:FreeEnergy} are sufficient to fully align the domain walls. As the film becomes thicker and more similar to a bulk crystal, more domain-wall sections align perpendicular to the z-direction. The ratio of perpendicular sections rises significantly more slowly for systems that include electrostatic energy contributions because of the energy penalty from charged domain walls.
Because the phase-field model is a freely scalable model, this crossover effect from thin film to bulk may be observed at different length scales in a real lab material compared to the scale of our simulation.

\section{Discussion}

    \label{sec:discussion}
    
In Figs.~\ref{fig:RealSpaceImages} to \ref{fig:ThickFilms}, we have shown that there are three effects that are caused by the depolarizing field exerted by bound charges at surfaces and interfaces of thin films. (i) The averaged magnitude of the polarization is lowered, (ii) the polarization declines away from the domain walls with a dip in polarization in the approximate center of the domain, and (iii) a flat domain size distribution and a large average domain size for simulations where the depolarizing field is ignored contrasts with a distribution with a clear peak and a smaller average domain size for simulations that include the depolarizing field. Furthermore, due to the electric field generated by bound charges at domain walls, domain walls have a stronger tendency to align parallel to the z-axis if the electrostatic energy is considered, as shown in Figs.~\ref{fig:ThickFilms} and \ref{fig:ThickFilmsQuantitative}.

Note that modification of the domain-size shape and distribution is even more surprising considering that the ferroelectricity in hexagonal manganites is not only improper, but also exhibits topological defects. Despite this two-fold opposition to the action of the depolarizing field, it manifests itself with unprecedented clarity in the Fourier-transformed intensity and cross-correlation.
Our work shows that the depolarizing field can have a significant effect on the domain structure of improper ferroelectrics, even if topological defects are present, and that, in general, it is not appropriate to ignore it.

Further, the depolarizing field lowers but does not fully suppress the polarization of the system.
This is characteristic for improper ferroelectrics and has been shown for a simplified Landau expansion of YMnO$_3$ in Sai et al. \cite{sai_absence_2009}. This result can be extended to Eq.~\ref{eq:FreeEnergy}, that is, the full Landau expansion of hexagonal manganites, and therefore our simulations also show this characteristic behavior.
This contrasts with proper ferroelectrics, where a strong depolarizing field can prevent a polarization from emerging in the first place.

As mentioned in Section \ref{Landau}, our choice of $\varepsilon_{\rm b}=1$ enhances the effect of the depolarizing field in our simulations without changing these effects qualitatively. In the supplemental material, we show a series of simulations between the physical limit of $\varepsilon_{\rm b}=1$ and the experimental value of $\varepsilon_{\rm b}=8.9$, and it is apparent how the influence of the depolarizing-field decreases with $\varepsilon_{\rm b}$. While it may be argued that the depolarizing-field correction may eventually become negligible, we point out that even a small correction is likely to have an effect on domain formation in the first place, as it occurs at the point of instability when crossing the Curie temperature.

\section{Conclusion}

    In this work, we have shown that the depolarizing field has a significant effect on the domain pattern of both the primary lattice-distortive and the secondary ferroelectric order in thin film hexagonal manganites. We find that the average magnitude of the polarization in the domains is reduced, the system exhibits local variations in polarization, and there is a significant change in the domain-size distribution and the average domain size. Furthermore, the energy penalty of charged domain walls causes the domain walls to align parallel to the direction of the polarization. 
    
    The emergence of these effects shows that 
    although the primary order and the topological defects are still the main driving force behind the domain structure, the effect of the depolarizing field is significant and, in general, non-negligible. We have also shown these effects are inconspicuous and only reveal themselves clearly in reciprocal space, in contrast to the domain structure of the topological defects which is apparent in real space. This may explain why these effects have not been considered in previous work.

    Control of ferroelectric domains is a requirement for thin film ferroelectric devices, which is parcitularly challenging in improper ferroelectrics because the ferroelectric domain structure is pinned to the primary order. Topological defects make control even more challenging. Known avenues of control have been pinning due to local interface effects \cite{bortis_manipulation_2022} and application of strain \cite{xue_strain-induced_2017, sandvik_pressure_2023}. Here, we have shown that the depolarizing field, despite directly acting on the secondary ferroelectric order only, also couples to the primary order and has a significant, global effect on the domain pattern of the primary order. Unlike strain, the depolarizing field preserves the isotropy in the ab-plane and does not affect the topological defects. We speculate that in a strained thin film, the depolarizing field may affect the stripe domains in qualitatively the same way as in isotropic domains of an unstrained sample and may change the width and onset of the domain stripes. This opens an additional avenue for control of domain patterns and the prospect of thin film devices of improper ferroelectrics.

\begin{acknowledgments}
We acknowledge useful discussions with Nicola Spaldin, Morgan Trassin, and Quintin Meier. We further thank Martin Lilienblum for providing the piezoresponse-force-microscopy images shown in the supplementary material. This work was funded by the Swiss National Science Foundation (SNSF) through grant numbers 200021\_178825 and 200021\_215423.
\end{acknowledgments}

\appendix

\section{Detailed Numerical Methods}
\label{app:NumericalMethodsAppendix}

\textit{Terms and parameters of the Landau expansion.} We use the Landau expansion as given by Artyukhin et al. \cite{artyukhin_landau_2014}. The structural mode of the system is described by the terms

\begin{equation}
F_\text{structural} = \frac{a}{2} Q^2 + \frac{b}{4} Q^4 + \frac{Q^6}{6} (c + c' \text{cos} \, 6 \Phi),
\end{equation}

\noindent where $Q$ and $\Phi$ are the amplitude and azimuthal angle of the structural mode respectively, and $a$, $b$, $c$, and $c'$ are parameters of the Landau expansion. Here, we use the values $a = -2.626 \, \text{eV} \text{\AA}^{-2}$, $b = 3.375 \, \text{eV} \text{\AA}^{-4}$, $c = 0.117 \, \text{eV} \text{\AA}^{-6}$, and $c' = 0.108 \, \text{eV} \text{\AA}^{-3}$ \cite{artyukhin_landau_2014}. 

The structural order couples to the ferroelectic order. This is described by the Landau terms

\begin{equation}
F_{\text{Ferroelectric}} = -g Q^3 \mathcal{P} \, \text{cos} 3 \Phi + \frac{g'}{2} Q^2 \mathcal{P}^2 + \frac{a_\mathcal{P}}{2} \mathcal{P}^2
\end{equation}

\noindent with parameters $g = 1.945 \, \text{eV} \, \text{\AA}^{-4}$, $g' = 9.931 \, \text{eV} \text{\AA}^{-4}$, and $a_\mathcal{P} = 0.866 \, \text{eV} \, \text{\AA}^{-2}$ \cite{artyukhin_landau_2014}. Because $a_\mathcal{P}$ is positive, the polar mode $\mathcal{P}$ cannot emerge on its own in the absence of structural order. The structural order causes the ferroelectric order, so the latter is improper.

The gradient terms of the system are given by

\begin{equation}\begin{split}
   F_{\text{gradient}} = \frac{1}{2} \sum  \limits_{i = x,y,z} \big [ s_Q^i \big ( \partial_i Q \partial_i Q + Q^2 \partial_i \Phi \partial_i \Phi \big ) & \\
    + s_\mathcal{P}^i \, \partial_i \mathcal{P} \partial_i \mathcal{P} \big ] & , 
\end{split}
\end{equation}

\noindent with parameters $s_\text{Q}^\text{z} = 15.40 \, \text{eV}$, $s_\text{Q}^\text{x} = 5.14 \, \text{eV}$, $s_\mathcal{P}^\text{z} = 50.70 \, \text{eV}$, and $ s_\mathcal{P}^\text{x} = 8.88 \text{eV}$ \cite{artyukhin_landau_2014} Here, $s_\mathcal{P}^\text{x}$  has been chosen differently from \cite{artyukhin_landau_2014} in order to ensure stability of the system. In \cite{xue_topological_2018}, this choice has been validated to give physical results, consistent with the parameters in \cite{artyukhin_landau_2014}. The gradient term corresponds to a penalty for non-homogeneous systems and therefore is responsible for the domain-wall energy.

Finally, a term

\begin{equation}
    F_\text{electrostatic} = -P E
\end{equation}

\noindent is added, where $P = \mathcal{P} \cdot 9.031 \,  e / 365.15 \, \text{\AA}^3 $ is the polarization of the system \cite{artyukhin_landau_2014} and $e$ is the elemental charge. The electric field $E$ is computed via the Gauss law. We assume that our system is a perfect insulator, and hence the total charge density $\rho$ is given by bound charges only according to $\rho = - \nabla P$. From the charge density, the electrostatic potential is computed via the Poisson equation $\Delta \varphi = - \rho / \varepsilon_{\rm b} \varepsilon_0$. The background dielectric constant contains all dielectric contributions that are not caused by the spontaneous polarization $P_\mathrm{s}$, that is, electronic responses and responses from all normal modes except $\Gamma^{-}_{2}$ \cite{smabraten_charged_2018}. Finally, from the electrostatic potential, the electric field can be obtained via $E = - \nabla \varphi$.

\textit{Computational details.} We obtain the Ginzburg-Landau equations by transforming Eq.~\eqref{eq:FreeEnergy} to Cartesian coordinates and computing its variational derivative. We choose the parameter $L = 1$. We assume open boundary conditions along the hexagonal axis and periodic boundary conditions in the plane perpendicular to the hexagonal axis. In particular, we assume $E = 0$ outside the thin film, which corresponds to open-electrostatic or open-circuit boundary conditions.

We initialize the order parameters $Q_x = Q \, \text{cos}\Phi$, $Q_y = Q \, \text{sin}\Phi$, and $\mathcal{P}$ with a random, uniform distribution on the interval [-0.1, 0.1]. Qualitatively, the domain pattern is independent of the initial random configuration. Quantitatively, however, the specific domain pattern depends on the initial configuration.
The Ginzburg-Landau equations are then integrated in a finite difference scheme with a Runge-Kutta 4 integrator. For the main results in Figs.~\ref{fig:RealSpaceImages} to \ref{fig:PairCorrelation}, we simulate a system of size $n_x \times n_y \times n_z = 128 \times 128 \times 3$ with lattice spacing $\Delta x = \Delta y = 0.1 \, \text{nm}$ and $\Delta z = 0.3 \, \text{nm}$. Although phase-field simulations are a freely scalable model, it is still sensible to choose a physically meaningful scale for the simulations. Here, the thickness of the simulated film in $z$-direction has been chosen to approximately correspond to a single monolayer, which is the limit of the physically meaningful range. With this choice, all domain walls are parallel to the z-direction, and there are no charged domain walls. Bound charges only accumulate at the surfaces of the thin films and all electrostatic effects stem from the depolarizing field. We use a time step of $\Delta t = 5 \cdot 10^{-4}$. The background dielectric constant of the material $\varepsilon_{\rm b}=1$ is chosen as described in the main text and the supplemental material. The system was iterated for $n = 10^{4}$ time steps. 
The Poisson equation to obtain the electrostatic potential is solved using a custom V-multigrid solver using Jacobi iterations. Simulations of thick films in Fig.~\ref{fig:ThickFilms} have been performed with system of size $n_x \times n_y \times n_z = 128 \times 128 \times 15$ with lattice spacing $\Delta x = \Delta y = 0.2 \, \text{nm}$ and $\Delta z = 0.3 \, \text{nm}$ and with a time step of $\Delta t = 5 \cdot 10^{-4}$. Simulations including the electrostatic energy were iterated for $n = 10^{5}$ time steps, and $\varepsilon_{\rm b} = 1$. The system that ignores electrostatic interactions was iterated for $n = 4.7 \cdot 10^4$ time steps to obtain domains of similar size.
For the quantitative analysis of thick films in Fig.~\ref{fig:ThickFilmsQuantitative}, simulations with system of size $n_x \times n_y = 128 \times 128$ with variable $n_z$ have been performed, with lattice spacing $\Delta x = \Delta y = 0.2 \, \text{nm}$ and $\Delta z = 0.3 \, \text{nm}$. For the thinnest films, we chose $n_z = 3$, which approximately corresponds to the thickness of a single monolayer. The time step has been set to $\Delta t = 5 \cdot 10^{-4}$. The system was iterated for $n = 10^{5}$ time steps for simulations that include and that ignore the electrostatic interaction. For the simulations with electrostatic interaction, we have set $\varepsilon_{\rm b} = 1$.

\textit{Data analysis.} We compute the Fourier-transformed intensity of the system by first taking a slice of the computational mesh of the thin film at $z = 0$.
We first subtract the average of the polarization from the system and then compute the two-dimensional Fourier transform of the domain pattern.

\begin{equation}
   F(\mathbf k) = \mathcal{F}[ P(\mathbf r)-P_{\text{av}}]
\end{equation}

\noindent Here, $F(\mathbf k)$ corresponds to the structure factor, $\mathcal{F}$ denotes the Fourier transform and $P_{\text{av}}$ denotes the average polarization. We square the absolute values of the structure factor to obtain the Fourier-transformed intensity of the system.

\begin{equation}
   I(\mathbf k) = |F(\mathbf k)|^2
\end{equation}

\noindent Fourier-transformed intensities of twenty simulations are combined to obtain better statistics.
Furthermore, we perform data augmentation by applying a 90° rotation, a mirror operation, and a combined 90° rotation and mirror operation.
The pair correlation is obtained by then transforming the Fourier-transformed intensity back to real space.

\begin{equation}
   A(\mathbf r) = \mathcal{F}^{-1}\left[I(\mathbf k)\right] = \mathcal{F}^{-1}\left[|F(\mathbf k)|^2 \right] 
\end{equation}

\noindent As shown in the supplemental material, the full-width half-maximum of the pair correlation is proportional to the average domain size of the system.

We compute the domain-wall section ratio as in \cite{bortis_manipulation_2022}. The number of domain walls $N_\text{w}$ is computed by counting the number of sign changes of the polar mode $\mathcal{P}(x,y,z)$ along the x-, y-, and z-axis. The ration of domain walls $\rho_z$ in z-direction is then computed with

\begin{equation}
\begin{split}
         \rho_z = \frac{1}{N_\text{w}} \sum \limits_{x,y = 1}^{L_x,L_y} \sum \limits_{z = 1}^{L_z - 1} |  H(\mathcal{P}(x,y,z+1)) & \\
         - H(\mathcal{P}(x,y,z)) &|,       
\end{split}
\end{equation}

where $H$ is the Heaviside step function.

%
%

\end{document}


\preprint{APS/123-QED}

\title{Supplemental Material: Effect of the depolarizing field on the domain structure of an improper ferroelectric}%

\author{Aaron Merlin Müller}
 \affiliation{%
 Department of Materials, ETH Zurich, 8093 Zurich, Switzerland
}%
\author{Amadé Bortis}%
\affiliation{%
 Department of Materials, ETH Zurich, 8093 Zurich, Switzerland
}%

\author{Arkadiy Simonov}%
\affiliation{%
 Department of Materials, ETH Zurich, 8093 Zurich, Switzerland
}%

\author{Manfred Fiebig}%
\email{manfred.fiebig@mat.ethz.ch}
\affiliation{%
 Department of Materials, ETH Zurich, 8093 Zurich, Switzerland
}%

\author{Thomas Lottermoser}%
\affiliation{%
 Department of Materials, ETH Zurich, 8093 Zurich, Switzerland
}%

\date{\today}%

\maketitle

\section{Comparing the pair correlation for different background dielectric constants}
We perform the simulations for the pair correlation as shown in Fig.~4 of the main text for background dielectric constants of $\varepsilon_{\rm b} = 1,2,4,8$, and $8.9$ (where the latter is the computed value for YMnO$_3$ \cite{smabraten_charged_2018}), and for the case with no electrostatic interaction. With the exception of the varying background dielectric constant, we use the same parameters as in the main text. We combine simulations with 20 different initial values to achieve better statistics. To further improve our statistics, we perform data augmentation by performing a 90° rotation, a mirror operation, and a combined 90° rotation and mirror operation. 

\begin{figure}[h]
    \centering
    \includegraphics[width=0.9\textwidth]{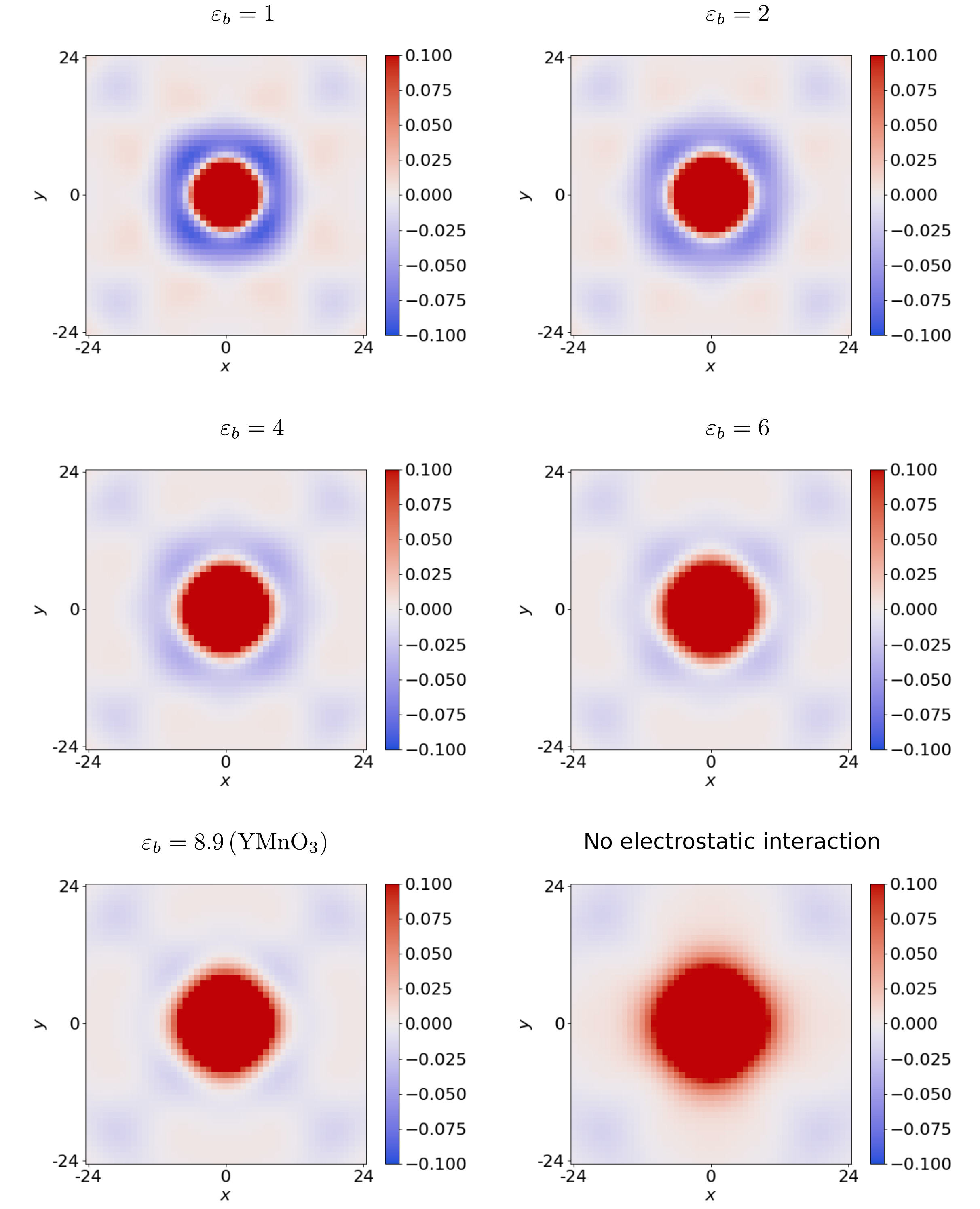}
    \caption{Pair-correlations for $\varepsilon_{\rm b} = 1, 2, 4, 6,$ and $8.9$, and for simulations without electrostatic interaction. This visualization shows the radius of a disk of positive correlations becoming gradually larger for simulations that include the electrostatic interaction as the background dielectric constant increases, which corresponds to an increase in average domain size. Furthermore, a characteristic ring for negative correlations is only visible in simulations that include electrostatic interaction.}
    \label{fig:pair_correlation}
\end{figure}

\begin{figure}[h]
    \centering
    \includegraphics[width=0.9\textwidth]{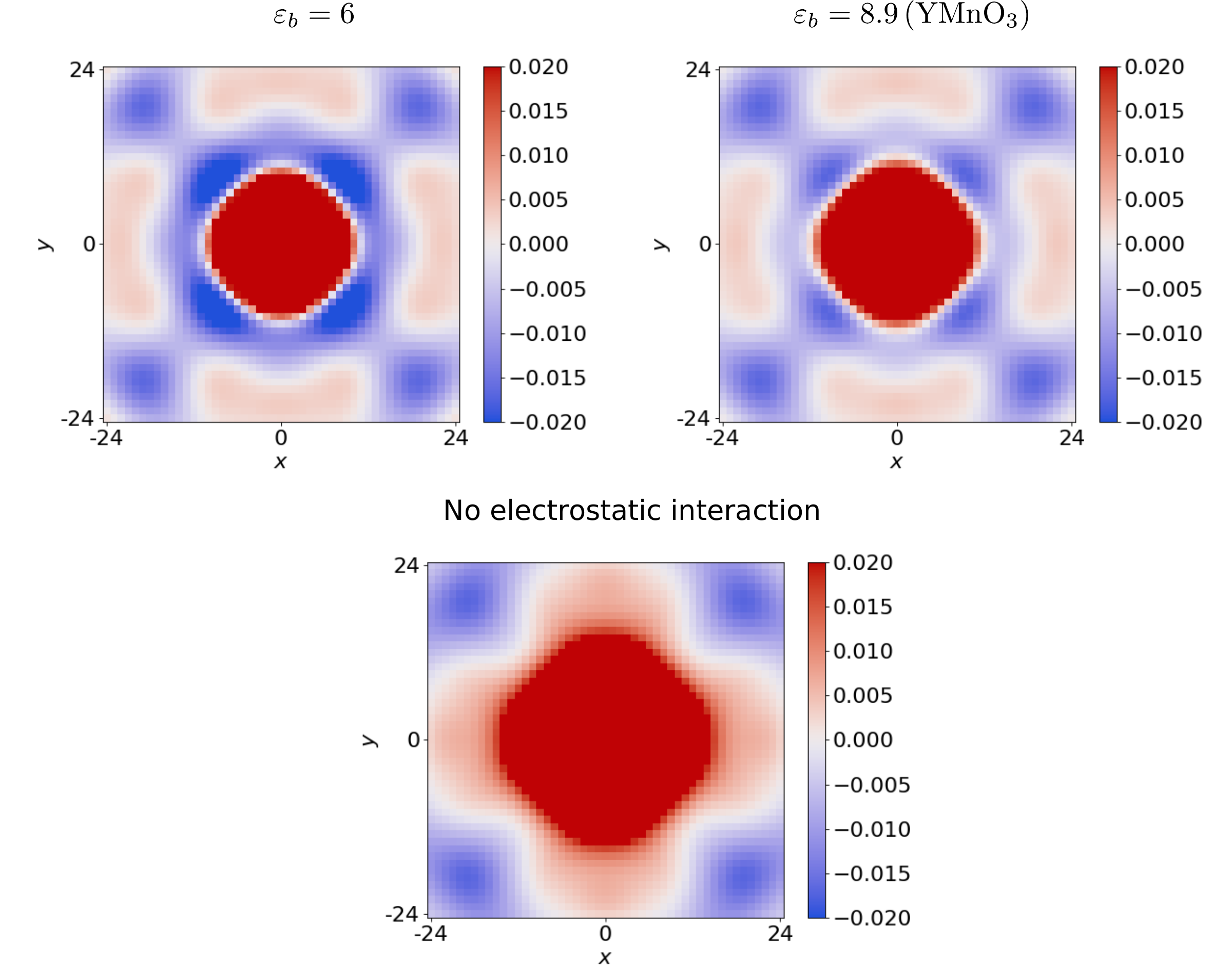}
    \caption{The same data as in Fig.~\ref{fig:pair_correlation} with different color-bar limits for $\varepsilon_{\rm b} = 6$, $\varepsilon_{\rm b} = 8.9$, and a simulation without electrostatic interaction. In this visualization, it is particularly pronounced that the ring of negative correlations is present for simulations with electrostatic interaction, whereas it is absent for simulations without electrostatic interaction, indicating a broader domain size distribution.}
    \label{fig:pair_correlation_additional}
\end{figure}

In Fig.~\ref{fig:pair_correlation}, we visualize our result. In Fig.~\ref{fig:pair_correlation_additional}, we visualize the same data as in Fig.~\ref{fig:pair_correlation} but with smaller color-bar limits. These visualizations suggest that although the domain size distribution changes in magnitude for increasing values of $\varepsilon_{\rm b}$, the qualitative result is the same for all simulations that include the electrostatic interaction. In contrast, the case without the electrostatic interaction does not show the characteristic ring of negative correlation and a larger central peak of positive correlation, indicating a broader domain size distribution and a larger average domain size.

\begin{figure}[h]
    \centering
    \includegraphics[width=0.7\textwidth]{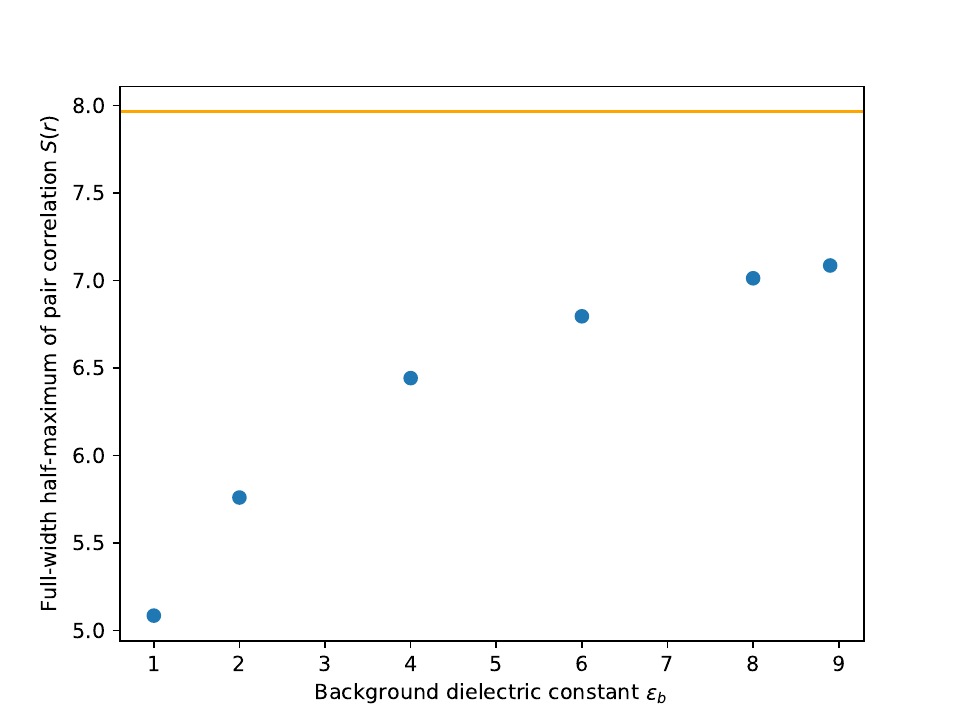}
    \caption{ Full-width half-maximum of the inner peak in Fig.~\ref{fig:pair_correlation} for varying background dielectric constants and the case without electrostatic interaction (orange line). This shows a difference in average domain size for background dieletric constants up to $\varepsilon_{\rm b} = 8.9$. }
    \label{fig:inner_disk_radius}
\end{figure}

 The effect of the electrostatic interaction on the average domain size is further illustrated in Fig~\ref{fig:inner_disk_radius}, where we plot the full-width half-maximum of the central peak of positive correlation. In the next section, we show that the full-width half-maximum of the central peak correlates with the average domain size. Hence, the electrostatic interaction has a significant effect on the average domain size even for $\varepsilon_{\rm b} = 8.9$

\clearpage

\section{Polarization profile of different background dielectric constants}

\begin{figure}[h!]
    \centering
    \includegraphics[width=0.7\textwidth]{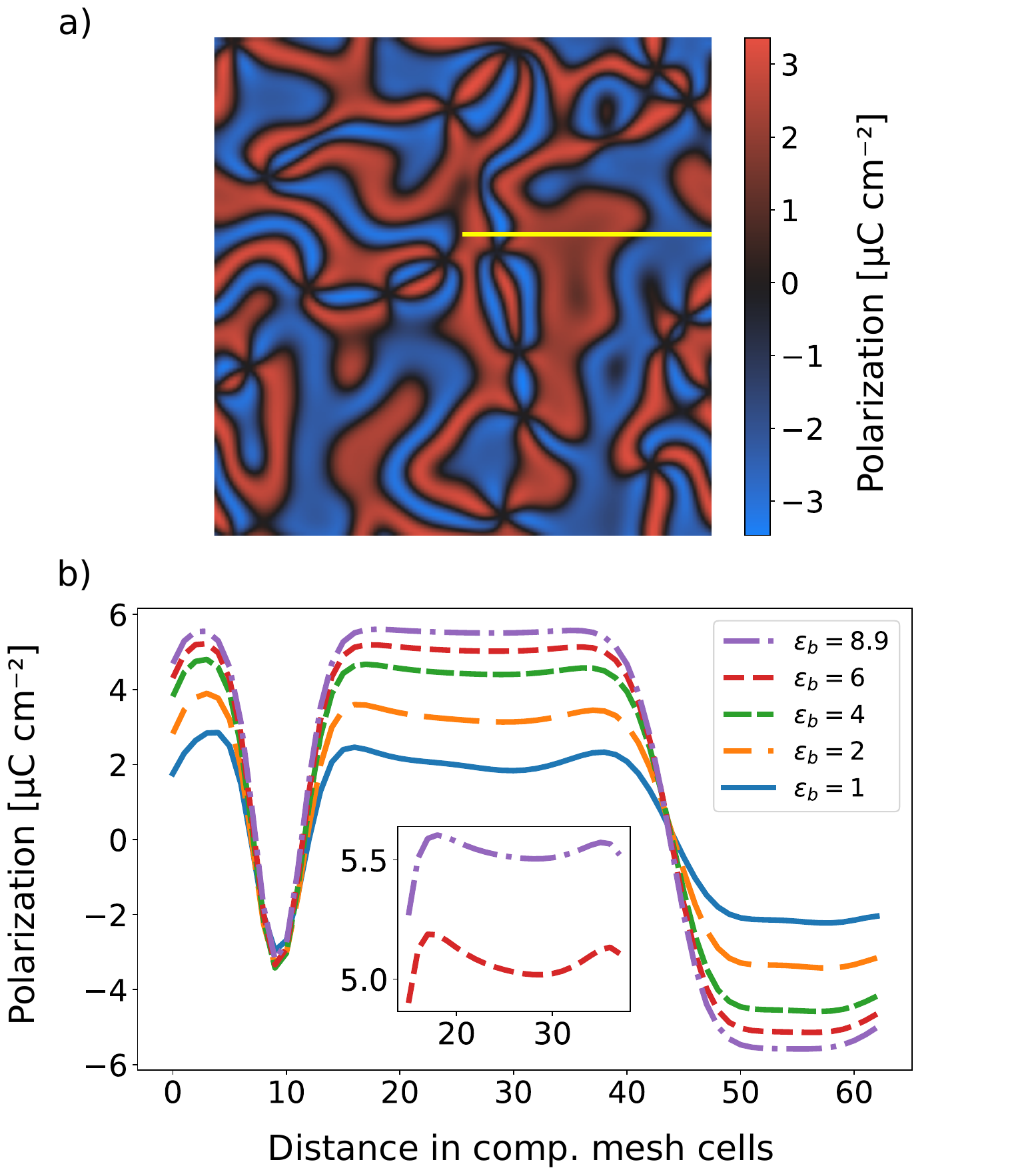}
    \caption{(a) Visualization of the domain pattern of a system starting from the same initial field as in Fig.~2b in the main text, with background dielectric constant $\varepsilon_{\rm b} = 1$. (b) Polarization profile along the yellow section in (a) for varying background dielectric constants. The average magnitude of the polarization is larger for larger background dielectric constants. The insets show that the dip along the polarization profile is present for all background dielectric constants. This shows that the results are qualitatively the same for realistic background dielectric constants and $\varepsilon_{\rm b} = 1$, as chosen in the main text.}
    \label{fig:polarization_profile_supp}
\end{figure}

We repeat the simulations of Fig.~2b in the main text, using the same parameters and initial fields but with varying background dielectric constants of $\varepsilon_{\rm b} = 1,2,4$, and $8.9$. We show in Fig.~\ref{fig:polarization_profile_supp} that as the background dielectric constant increases, the term including the electrostatic interaction becomes smaller, resulting in a higher absolute polarization density and decreasing the dip in the polarization profile. Qualitatively, however, the effect remains the same, even if it becomes quantitatively weaker. This justifies the choice of $\varepsilon_{\rm b} = 1$ in the main text, which enhances the effects of the electrostatic interaction and permits a reduction of the size of our computational mesh, keeping the computational requirements feasible.

\section{Estimation of domain size}

The domain size was estimated from the pair-correlation function of ferroelectric domain pattern. This method is very simple and efficient, since it can be calculated using the fast-Fourier transform. The calculation works as follows. We start from a map of ferroelectric polarisation $P(\mathbf r)$, first we subtract the average polarization and then use a Fourier transform to calculate the structure factor:

$$ F(\mathbf k) = \mathcal{F}[ P(\mathbf r)-P_{\text{av}}] $$

\noindent where $\mathcal{F}$ denotes the Fourier transform. The pair-correlation function of the ferroelectric order parameter can then be calculated using the inverse Fourier transform:

$$ A(\mathbf r) = \mathcal{F}^{-1}\left[|F(\mathbf k)|^2 \right] $$

\noindent The pair-correlation function has a sharp peak in the center which quickly decays to zero. The radius of the central peak is proportional to the domain size. We define the domain size $d$ as the average value of the full width at half maximum of the pair-correlation function:

$$ d = \text{FWHM}[A(\mathbf r)] $$

\begin{figure}
    \centering
    \includegraphics[width=0.95\linewidth]{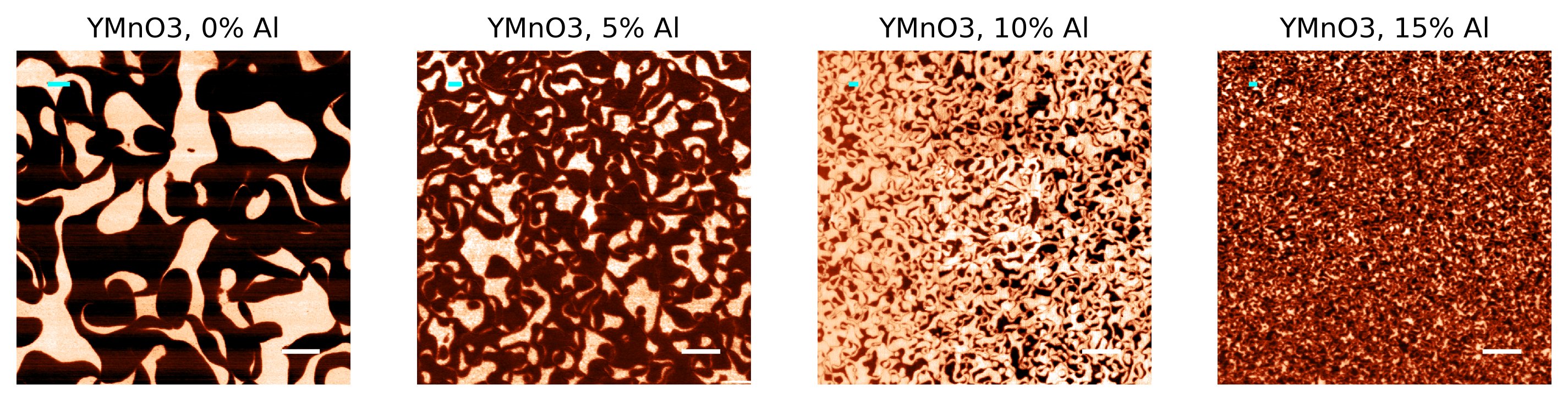}
    \caption{Domain pattern from Al doped YMnO$_3$ single crystals. The white scale bar in the lower-right corner represents 1 µm, while the cyan mark in the upper-left corner corresponds to the domain size estimated using the described method.}
    \label{fig:domain_patterns_experimental}
\end{figure}

\begin{figure}
    \centering
    \includegraphics[width=0.5\linewidth]{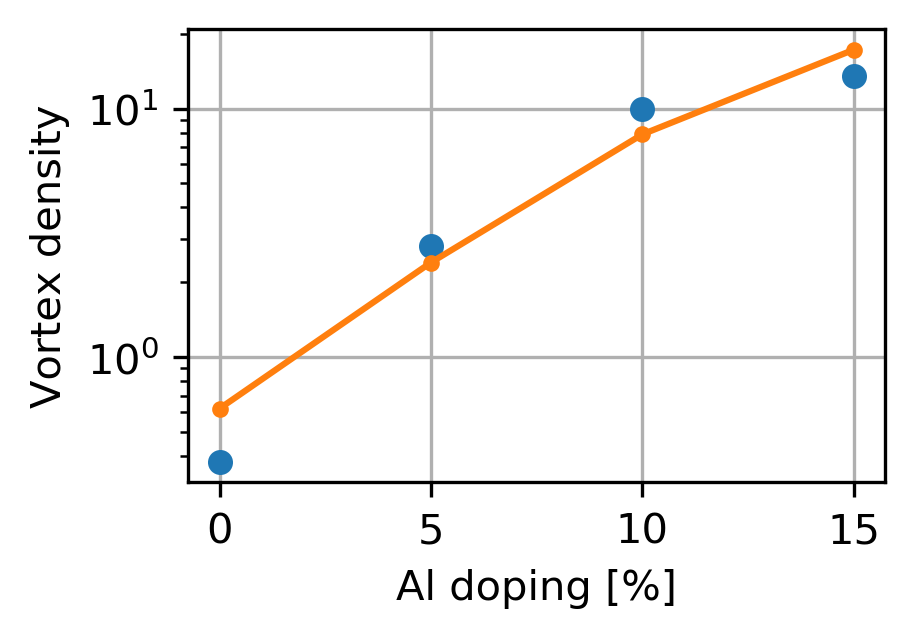}
    \caption{Comparison of different domain size estimation methods. Blue markers correspond to measured vortex density expressed in number of vortices per 1 µm$^2$. The data presented in the orange line is equal to $1/(6d^2)$ where $d$ is the average domain size estimated from the pair-correlation function.}
    \label{fig:Domain_Size_Calibration}
\end{figure}

\noindent This method was tested extensively in our previous work \cite{giraldoMagnetoelectricDomainEngineering2024a}, and was found to be equally efficient as more more laborious methods, such as measuring the vortex density, \emph{i.e.} by measuring the number of topological defects per unit area. Figure \ref{fig:domain_patterns_experimental} present the experimental domain pattern from Al doped YMnO$_3$, and the Figure \ref{fig:Domain_Size_Calibration} presents a comparison of the domain-size estimation using the pair-correlation method versus an estimation by measuring the vortex density \cite{giraldoMagnetoelectricDomainEngineering2024a}.

%
%